\documentclass[aps,prl,twocolumn,showpacs,superscriptaddress,groupedaddress,amsmath,amssymb]{revtex4-2}
\usepackage{graphicx}
\usepackage{latexsym}
\usepackage{bm} 
\usepackage{color}
\usepackage{epsfig}
\usepackage{multirow}
\usepackage{xcolor}
\usepackage{colortbl}
\usepackage{hhline}
\usepackage{simplewick}
\usepackage{soul}

\usepackage{graphicx}
\usepackage[colorlinks=true,linkcolor=blue,citecolor=blue,urlcolor=blue]{hyperref}
\usepackage{bbold}
\usepackage{gensymb}

\AtBeginDocument{%
    \newwrite\bibnotes
    \def\bibnotesext{Notes.bib}
    \immediate\openout\bibnotes=\jobname\bibnotesext
    \immediate\write\bibnotes{@CONTROL{REVTEX42Control}}
    \immediate\write\bibnotes{@CONTROL{%
    apsrev42Control,author="08",editor="1",pages="0",title="0",year="1"}}
     \if@filesw
     \immediate\write\@auxout{\string\citation{apsrev42Control}}%
    \fi
}%

\def \mbf {\mathbf}

\newcommand{\comment}[1]{}

\definecolor{mygreen}{rgb}{0, 0.7, 0}

\begin{document}

\title{Sublattice polarization from destructive interference on common lattices}

\author{Yu-Ping Lin}
\affiliation{Department of Physics, University of California, Berkeley, California 94720, USA}

\date{\today}

\begin{abstract}
We show that sublattice-polarized states (SLPSs) appear ubiquitously on the common lattices. We first establish the destructive-interference (DI) scenario for the SLPSs, which is systematized by a point-group-symmetry interpretation. The examples on common one-, two-, and three-dimensional lattices are then demonstrated. We also deduce the symmetry-protected robustness of SLPSs against further-neighbor hoppings. Moreover, the DI scenario can be generalized to the multi-SLP. The important effects on interaction-driven phases are studied by Hartree-Fock analysis.
\end{abstract}

\maketitle

The understanding of wavefunction structures has become an important pillar in modern condensed matter physics. Extensive research has developed in the momentum-space framework, where the essential roles of topology \cite{chiu16rmp} and geometry \cite{rossi21co} are uncovered. Recent studies further reveal the importance of wavefunction structures to the interaction-driven phases. A prominent example is the kagome lattice at its middle-band Van Hove singularity, which is relevant to the kagome metals $A$V$_3$Sb$_5$ with $A=$ K, Ru, Cs \cite{ortiz19prm,ortiz20prl,jiang21nm,zhao21n} and FeGe \cite{teng22n,yin22prl,teng23np}. Due to strong Fermi-surface nesting, one may expect the electronic repulsion to drive spin-density waves or fluctuation-induced superconductivity \cite{honerkamp03prb,gneist23prb,honerkamp08prl,nandkishore12np,kiesel12prl,wang12prb}. However, the Fermi surface exhibits an intriguing {\it sublattice polarization} ({\it SLP}, or sublattice interference) \cite{kiesel12prb}, where the wavefunction at each saddle point solely resides in a sublattice even under third-neighbor hoppings \cite{wu23prb}. The SLP strongly obstructs the nesting effect and favors intra-unit-cell (IUC) orders at weak coupling. Indeed, functional-renormalization-group (FRG) analyses identify ferromagnetism (FM) and IUC charge-density modulations (CDMs) at weak coupling, together with other interesting phases at moderate coupling \cite{kiesel13prl,wang13prb,profe24ax,schwemmer23ax}. Similar SLP effects on interaction-driven phases are recently studied on the honeycomb lattice \cite{castro23prl}. Despite the extensive studies of interaction-driven phases on these specific lattices, the origin and generality of SLP remain elusive. Understanding the SLP can advance the search for exotic correlated phases across materials, and further boost the quantum technology through sublattice control.

The wavefunction structures have also been investigated in the real-space framework. In particular, recent works recognize the {\it destructive interference (DI)} as the source of flat bands on various lattices \cite{sutherland86prb,mielke91jpa,tasaki92prl,bergman08prb,moralesinostroza16pra,maimaiti17prb,rhim19prb,morfonios21prb,graf21prb,neves24npjcm}. Under frustrated hoppings, some loop wavefunctions can fail to spread and remain invariant. These compact localized and noncontractible loop states serve as the dispersionless eigenstates and form the flat bands. Since the SLP shows a similar failure of spreading under hoppings, it is natural to ask: {\it can SLP also be understood from DI?}

In this {\it Letter}, we confirm the answer by examining the DI between sublattices. Remarkably, this approach establishes the fundamental understanding of {\it sublattice-polarized states (SLPSs)} and applies generally to the common lattices. We first set up the general principle of finding SLPSs from DI. This scenario is reinforced by a point-group-symmetry interpretation, which enables systematic search for SLPSs and justifies their robustness against further-neighbor hoppings. We then demonstrate the examples on common one-, two-, and three-dimensional (1D, 2D, and 3D) lattices. The generality of DI further allows us to consider the multi-SLP (MSLP). Finally, we study the important effects of SLP on interaction-driven phases by Hartree-Fock analysis.

\textit{SLPS.---}We begin with the DI scenario for the SLPSs. Our analysis considers the tight-binding model
\begin{equation}
H=-\sum_{ii'}\sum_{\tau\tau'}t_{ii'\tau\tau'}c_{i\tau}^\dagger c_{i'\tau'}
\end{equation}
on multisublattice lattices. Here $c_{i\tau}^{(\dagger)}$ annihilates (creates) a fermion in sublattice $\tau=0,1,2,\dots$ at Bravais-lattice site $i$. The hoppings $t_{ii'\tau\tau'}$ are defined by their ranges of action as onsite $t_0$, nearest-neighbor $t_1$, second-neighbor $t_2$, etc. As a starting point, we focus on the uniform nearest-neighbor-hopping model with $t_1=1$. For any two sublattices $\tau$ and $\tau'$, we define the nearest-neighbor connection index $N^{\tau\tau'}_1=N^{\tau'\tau}_1$, which is the number of nearest-neighbor $\tau$ ($\tau'$) sites for a $\tau'$ ($\tau$) site. Remarkably, we find SLPSs on a broad set of common lattices with $N^{\tau\tau'}_1\geq2$ or $=0$. Note that ``trivial" SLPSs always exist in an isolated sublattice $\tau$, where $N^{\tau\tau'}_1=0$ for any $\tau'\neq\tau$. We will focus on the nontrivial situations with $N^{\tau\tau'}_1\geq2$ for at least one $\tau'\neq\tau$.

\begin{figure}[t]
\centering
\includegraphics[scale=1]{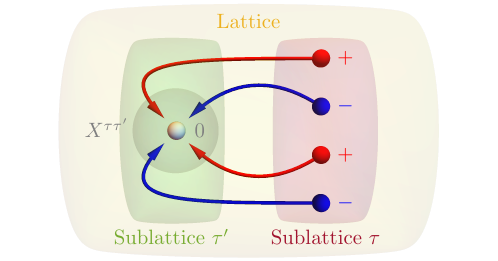}
\caption{\label{fig:di} DI scenario for SLP. An SLPS in a sublattice $\tau$ can fail to spread to the other sublattices $\tau'$ under DI. The DI at a site in $\tau'$ is governed by a point-group symmetry $X^{\tau\tau'}$.}
\end{figure}

When $N^{\tau\tau'}_1\geq2$, each site $i'\tau'$ is connected to $N^{\tau\tau'}_1$ sites $i\tau$. This structure allows for a DI from $\tau$ to $\tau'$ (Fig.~\ref{fig:di}). Consider an SLP wavefunction $\psi_{i\tau}$ in $\tau$ which sums to zero $\sum_i^{i'\tau',1}\psi_{i\tau}=0$ over the $N^{\tau\tau'}_1$ sites $i\tau$. Under hoppings, the wavefunction remains zero at the site $i'\tau'$. If such DI occurs at all sites $i'\tau'$ with $\tau'\neq\tau$, the wavefunction is an SLP eigenstate, which we call an SLPS.

Since a DI involves an $N^{\tau\tau'}_1$-site wavefunction structure, the SLPS should manifest a compatible enlarged periodicity. This condition implies a {\it nonzero} momentum $\mbf k_\text{SLPS}^\tau$, which usually puts the SLPS $(\tau,\mbf k_\text{SLPS}^\tau)$ at a high-symmetry point of the Brillouin zone (BZ). Note that DI may occur to multiple wavefunctions with different periodic structures, especially when $N^{\tau\tau'}_1\geq3$. In this case, SLPSs can appear at multiple high-symmetry points or even in extended domains. The energy $\epsilon_\text{SLPS}^\tau$ depends on the self connection of the sublattice. If the sublattice is self-disconnected, a common situation on most lattices, the SLPS vanishes under hopping at zero energy $\epsilon_\text{SLPS}^\tau=0$. Meanwhile, the energy becomes nonzero $\epsilon_\text{SLPS}^\tau\neq0$ when the sublattice is self-connected. SLPSs in different sublattices can appear at the same or different momenta. If different SLPSs appear at the same momentum $\mbf k_\text{SLPS}^\tau$ and energy $\epsilon^\tau_\text{SLPS}$, the degeneracy leads to a band crossing. This picture explains a set of band crossings transparently, and can be compared with related discussions of chiral symmetry \cite{dai24ax}. Pure SLPSs appear when the band crossings are gapped by sublattice potentials, sublattice-resolved hoppings, etc.

{\it Point-group-symmetry interpretation.---}Interestingly, SLPSs are intimately related to the point-group symmetry. Consider the structure of a sublattice $\tau$ around a site $i'\tau'$ with $\tau'\neq\tau$. The $N^{\tau\tau'}_1$ nearest-neighbor sites $i\tau$ obey a point-group symmetry $X^{\tau\tau'}$ at $i'\tau'$. Importantly, the DI condition $\sum_i^{i'\tau',1}\psi_{i\tau}=0$ corresponds to the nontrivial irreducible representations (irreps) \cite{bilbao11} $Y$'s with nonzero nearest-neighbor harmonics. The basis functions $w_y$'s of these irreps span a DI Hilbert space, where the elements characterize the eligible wavefunctions at $i\tau$. Appropriate choices of elements around all $i'\tau'$ then form an SLPS in $\tau$. From this interpretation, SLPSs can be identified systematically on general lattices.

Having established the DI scenario for the SLPSs, we now demonstrate its wide applicability to the common 1D, 2D, and 3D lattices (Supplementary Sec.~I).

\begin{figure}[t]
\centering
\includegraphics[scale=1]{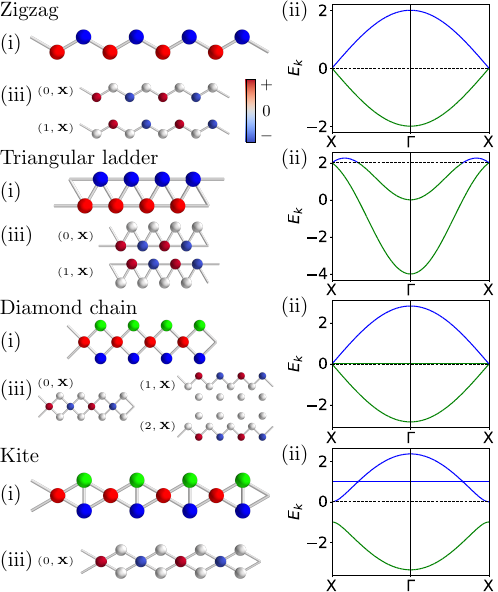}
\caption{\label{fig:1d} Examples of 1D lattices with SLPSs. For each lattice, we show (i) the lattice with sublattices marked by rainbow colors ($\tau=0,1,2,\dots$ from red to blue), (ii) the band structure with SLPS energy $\epsilon_\text{SLPS}^\tau$ indicated by dashed line, and (iii) the wavefunctions $\psi_{i\tau}$ of representative SLPSs $(\tau,\mbf k^\tau_\text{SLPS})$ with $\tau$-connected bonds illustrated.}
\end{figure}

\textit{1D lattices.---}We begin with the SLPSs on the 1D lattices (Fig.~\ref{fig:1d}). The connection index is at most $N^{\tau\tau'}_1=2$. Correspondingly, the relevant symmetry is an $x$-direction mirror symmetry equivalent to C$_\text{i}$. The nontrivial irrep $A_u$ has a basis function $w_1^2=(1/\sqrt2)(1,-1)$, which implies the staggered SLPS $\psi_{i\tau}=(-1)^i$ at the BZ edge $\mbf X$. We first consider the zigzag lattice with two sublattices and $N^{01}_1=2$. The SLPSs appear in both sublattices and form a nodal point. Similar SLPSs also appear on the triangular ladder lattice. However, the self connections of the sublattices push the SLPSs to nonzero energy $\epsilon_{\mbf X}=2$. We note in passing that the square-ladder lattice does not support SLPS due to $N^{01}_1=1$. On the other hand, the three-sublattice diamond-chain lattice \cite{aoki20jsnm} with $N^{01,02}_1=2$ also hosts SLPSs. Here DI applies to all sublattices, and the triple degeneracy leads to a three-band nodal point. There are also cases where DI does not apply to all sublattices. For example, on the kite lattice \cite{neves24npjcm} with $N^{01,02}_1=2$ and $N^{12}_1=1$, an SLPS appears singly in $\tau=0$.

\begin{figure*}[t]
\centering
\includegraphics[scale=1]{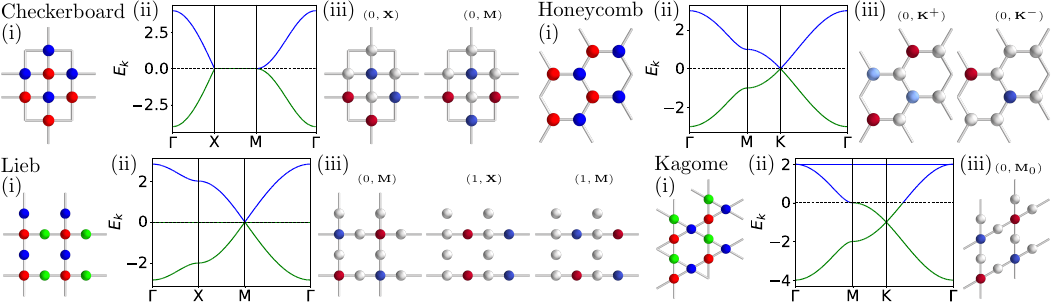}
\caption{\label{fig:2d} Examples of 2D lattices with SLPSs. For the honeycomb lattice, the labels $\mbf K^\pm$ correspond to $w^3_{1,2}$, respectively.}
\end{figure*}

\textit{2D lattices.---}We next explore the SLPSs on the 2D lattices (Fig.~\ref{fig:2d} and Supplementary Fig.~S1). Our first targets are the lattices on square Bravais lattice. The simplest example is the checkerboard lattice with two sublattices and $N^{01}_1=4$. Under the C$_{4v}$ symmetry, the nontrivial irreps $B_2$ and $E$ form a 3D DI Hilbert space. The basis functions $w_{n=1,2,3}^4=(1/2)(1,[-1]^{\delta_{1n}+1},[-1]^{\delta_{2n}+1},[-1]^{\delta_{3n}+1})$ imply the staggered SLPSs with $2\times1$, $1\times2$, and $\sqrt2\times\sqrt2$ periodicities, which sit at the BZ edge centers $\mbf X$, $\mbf Y$, and corner $\mbf M$, respectively. The SLP domain is further extended along the whole BZ boundary, where nodal lines occur under sublattice degeneracy. Extended SLP domains also appear on the three-sublattice Lieb lattice \cite{lieb89prl}. From the $N^{01,02}_1=2$ structures under the C$_\text{i}$ symmetry, the irrep $A_u$ indicates staggered SLPSs. Since the sublattice $\tau=0$ is connected to $\tau=1,2$ in $x$ and $y$ directions, an enforced $\sqrt2\times\sqrt2$ periodicity sets its SLPS at the BZ corner $\mbf M$ \cite{rhim19prb}. Meanwhile, the rest two sublattices $\tau=1,2$ are only connected to $\tau=0$ along $x$ or $y$ direction. This loose connection allows the $\tau=1,2$ SLPSs also at the BZ edge centers $\mbf X$ and $\mbf Y$, respectively, and further along the BZ edges $\mbf M$-$\mbf X$-$\mbf M$ and $\mbf M$-$\mbf Y$-$\mbf M$. At the BZ corner $\mbf M$, the triple degeneracy leads to a three-band nodal point.

SLPSs also arise on the lattices on triangular Bravais lattice. For the two-sublattice honeycomb lattice with $N^{01}_1=3$, the C$_{3v}$ symmetry implies a 2D DI Hilbert space from the nontrivial irrep $E$. A natural basis involves the basis functions $w^3_1=(1/\sqrt{6})(2,-1,-1)$ and $w^3_2=(1/\sqrt{2})(0,1,-1)$. The $\sqrt3\times\sqrt3$ periodicity assigns the SLPSs at the BZ corners $\mbf K$ and $\mbf K'=-\mbf K$. These SLPSs manifest their wavefunctions from $(w^3_1\pm iw^3_2)/\sqrt2$ and form nodal points \cite{castroneto09rmp} under sublattice degeneracy. Notably, the SLP gaps are observed under sublattice potentials in hexagonal boron nitride \cite{rubio94prb,blase94epl,watanabe04nm,topsakal09prb}, under strains in graphene \cite{schneider15prb,mao20n,lu23nc}, and under sublattice loop currents in theoretical Haldane model \cite{castro23prl}. Meanwhile, the SLPSs are known on the kagome lattice \cite{kiesel12prb}, where the BZ edge centers $\mbf M_{0,1,2}$ correspond to the three sublattices $\tau=0,1,2$. These SLPSs are naturally understood from our DI scenario. Given the $N^{\tau\tau'}_1=2$ structures, the C$_\text{i}$ symmetry implies staggered SLPSs from the nontrivial irrep $A_u$. The $2\times1$ periodicities then set the $\tau=0,1,2$ SLPSs at $\mbf M_{0,1,2}$.

\begin{figure*}[t]
\centering
\includegraphics[scale=1]{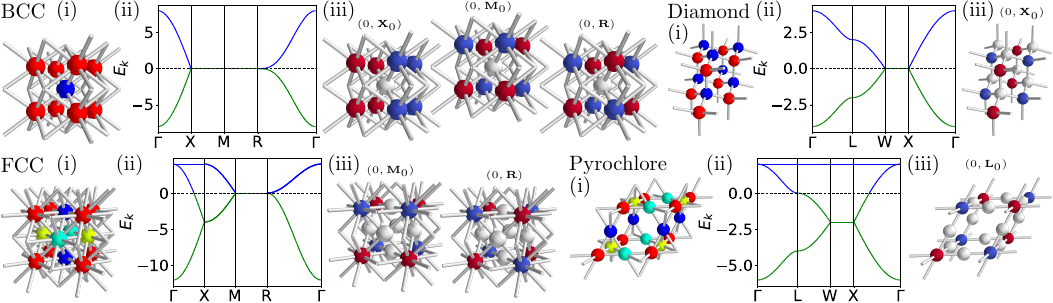}
\caption{\label{fig:3d} Examples of 3D lattices with SLPSs. The angles of view are chosen for clear illustration.}
\end{figure*}

\textit{3D lattices.---}We further go onto the 3D lattices (Fig.~\ref{fig:3d}, Supplementary Figs.~S2 and S3). For the 3D checkerboard and Lieb lattices, the SLPSs can be directly generalized from their 2D analogs. Here we discuss the SLPSs on other lattices. We first consider the lattices on simple cubic Bravais lattice. For the body-centered cubic (BCC) lattice with two sublattices, the large $N^{01}_1=8$ structures obey the $O_h$ symmetry. A 7D DI Hilbert space is determined by the irreps $A_{2u}$, $T_{2g}$, and $T_{1u}$, whose basis functions form the staggered SLPSs at the BZ corner $\mbf R$, edge centers $\mbf M_{0,1,2}$, and face centers $\mbf X_{0,1,2}$, respectively. The SLP domain covers the whole BZ boundary, where a nodal surface appears under sublattice degeneracy. We also consider the face-centered cubic (FCC) lattice with four sublattices. The $N^{\tau\tau'}_1=4$ structures obey the D$_{4h}$ symmetry, thereby manifesting a 3D DI Hilbert space from the irreps $B_{2g}$ and $E_u$. The SLP domain extends along all BZ edges, where the edge centers $\mbf M_{0,1,2}$ and corner $\mbf R$ manifest staggered SLPSs from the basis functions $w^4_{1,2,3}$. Under sublattice degeneracy, the SLP domain hosts four-band nodal lines.

Another common class of 3D lattices features the FCC Bravais lattice. One example is the diamond lattice with two sublattices. Under the T$_d$ symmetry, the $N^{01}_1=4$ structures host a 3D DI Hilbert space from the nontrivial irrep $T_2$. The basis functions $w^4_{1,2,3}$ constitute the staggered SLPSs at the BZ square-face centers $\mbf X_{0,1,2}$. The other elements further extend the SLP domain along the $\mbf X$-$\mbf X$ lines through the corners $\mbf W$'s. Under sublattice degeneracy, nodal lines occur along $\mbf X$-$\mbf W$. SLPSs also arise on the pyrochlore lattice with four sublattices. As a 3D analog of 2D kagome lattice, the pyrochlore lattice exhibits similar SLP high-symmetry points from the $N^{\tau\tau'}_1=2$ structures. With the irrep $A_u$ under the C$_\text{i}$ symmetry, each sublattice hosts its staggered SLPS at a BZ hexagon-face center $\mbf L_{0,1,2,3}$.

\textit{Robustness.---}Given the ubiquity of SLPSs on common lattices, it is important to understand their robustness beyond uniform nearest-neighbor hoppings. Remarkably, the robustness is usually symmetry-protected against uniform further-neighbor hoppings at each range. Consider two sublattices $\tau\neq\tau'$ on a lattice with $N^{\tau\tau'}_1\geq2$. Here we focus on the DI of an SLPS in $\tau$ to a site $i'\tau'$ under a point-group symmetry $X^{\tau\tau'}$. Crucially, the further neighbors $i\tau$ of $i'\tau'$ at each range form $N^{\tau\tau'}_1$ $X^{\tau\tau'}$-related groups, whose wavefunction sums correspond to an element in the DI Hilbert space. When further-neighbor hoppings are introduced uniformly at each range, the DI is still effective, and the SLPS remains an eigenstate. Therefore, SLPSs can be robust against further-neighbor hoppings. An alternative proof can be achieved by a mathematical induction (Supplementary Sec.~II). Note that the energies may change under new self connections of the sublattices. Exceptions to the robustness may occur when there are zero nearest-neighbor connection index $N^{\tau\tau'}_1=0$. In this case, the nonzero index $N^{\tau\tau'}_n$ at the shortest neighbor determines whether the SLPS remains robust. For example, the sublattices $\tau=1,2$ on the 2D Lieb lattice are disconnected at nearest neighbor $N^{12}_1=0$, but are highly connected at second neighbor $N^{12}_2=4$. The C$_{4v}$-symmetric structure still supports the original SLPSs. On the other hand, for the 1D diamond-chain lattice, the $\tau=1,2$ SLPSs are destroyed under the $N^{12}_2=1$ second-neighbor connection.

SLP can be suppressed when the hoppings become nonuniform at each range. The nonuniformity may result from, for example, beyond-$s$ orbitals, bond modulations, loop currents, or spin-orbit couplings. Nevertheless, SLP can still be stabilized if the system carries sublattice potentials or sublattice-resolved hoppings. Note that new SLPSs may also appear under nonuniform hoppings.

\textit{MSLP.---}When a lattice hosts more than two sublattices, MSLP may occur. The MSLP can also be understood from the DI scenario. Straightforwardly, the SLPSs in different sublattices can be combined into MSLPSs. Moreover, IUC DI becomes achievable for MSLPSs. If the IUC pattern is uniform over all unit cells, an MSLPS appears at the BZ center $\boldsymbol{\Gamma}$. Meanwhile, the modulated patterns lead to the MSLPSs at nonzero momenta. For example, with the wavefunction components $\psi_{i0}=0$ and $\psi_{i1}=-\psi_{i2}$ on the 1D diamond-chain and kite lattices, the bi-SLPSs form a flat band in the whole BZ. This wavefunction, together with its siblings under rotations, also give the bi-SLPSs along $\boldsymbol{\Gamma}$-$\mbf M$ lines on the 2D Lieb and kagome lattices at $\epsilon_{\boldsymbol{\Gamma}\text{-}\mbf M}=0$ and $2$, respectively. On the other hand, the increased connections allow more possible patterns under DI. For example, the wavefunction with $\psi_{i\tau_0}=0$ and $\psi_{i\tau_1}=\psi_{i\tau_2}$ can appear at $\mbf M_{0,1,2}$ and $\epsilon_{\mbf M}=-2$ on the 2D kagome lattice. Note that the MSLPSs can again be determined from the point-group symmetry. Taking the 2D Lieb and kagome lattices as examples, the bi-SLPSs can be obtained from the nontrivial irreps of C$_{4v}$ and C$_{2v}$ symmetries, respectively.

\begin{figure}[t]
\centering
\includegraphics[scale=1]{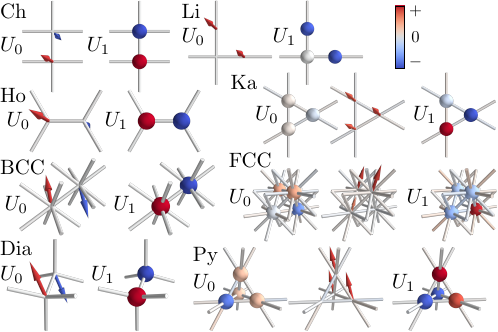}
\caption{\label{fig:hf} Interaction-driven ground states. For each lattice, we show the charge and/or spin patterns under pure onsite or nearest-neighbor repulsion $U_{0,1}>0$. The colors indicate the charge-density deviations from average or spin $z$ components on the sites and bonds.}
\end{figure}

\textit{Interaction-driven phases.---}The SLP has profound effects on the interaction-driven phases. Here we study the spin-$1/2$ fermions under Hubbard repulsions
\begin{equation}
H_\text{int}=\frac{1}{2}\sum_{ii'\tau\tau'}U_{ii'\tau\tau'}:n_{i\tau}n_{i'\tau'}:
\end{equation}
on the 2D and 3D lattices. The operator $n_{i\tau}=c_{i\tau}^\dagger c_{i\tau}$ represents the fermion density at $i\tau$, and $:\dots:$ indicates the normal ordering. The fillings are set by the zero SLPS energy $\epsilon_\text{SLPS}^\tau=0$. We compute the mean-field ground states by a Hartree-Fock analysis \cite{lin23ax} (Supplementary Sec.~III). Our focus is on the symmetry-breaking orders at the weakest couplings (Fig.~\ref{fig:hf}). Usually, symmetry-breaking orders develop at relevant momenta of the Fermi surface \cite{honerkamp03prb,gneist23prb,honerkamp08prl,wang12prb,ferhat14prb}. However, SLP can pin the orders at zero momentum. Under the onsite repulsion $U_0>0$, our computation finds IUC collinear spin orders generally. These orders act as spin-resolved sublattice potentials, thereby inducing opposite SLP splittings in the two spin branches. For the two-sublattice lattices, including 2D checkerboard, honeycomb, 3D BCC, and diamond lattices, the IUC antiferromagnetism (AFM) develops \cite{wu14prb,kiesel13prl,wang13prb,profe24ax}. Meanwhile, the other lattices develop IUC nonuniform FM with possible secondary IUC CDMs. On the other hand, the nearest-neighbor repulsion $U_1>0$ serves as an attractive SLP coupling $U_1n_{i\tau}n_{i'\tau'}\rightarrow -(U_1/2)(n_{i\tau}-n_{i'\tau'})^2$ and drives IUC CDMs \cite{wu14prb,kiesel13prl,wang13prb,profe24ax}. These orders generate sublattice potentials and induce SLP splittings. Finally, we note the possibility of more interesting phases at moderate coupling, where SLP strongly intertwines with other band-structure effects \cite{kiesel13prl,wang13prb,schwemmer23ax,profe24ax}.

\textit{Outlook.---}We have identified the DI as a fundamental origin of SLPSs. These SLPSs appear generally on the common lattices, remain robust against further-neighbor hoppings, and affect the interaction-driven phases profoundly. Our work opens an important gate toward the advanced understanding of wavefunction structures and their consequences. The information of potential platforms for sublattice control are beneficial to the quantum technology. There remains enormous uncharted territory on this research frontier, which extends and goes beyond the recent mainstream of kagome metals \cite{ortiz19prm,ortiz20prl,jiang21nm,zhao21n,teng22n,yin22prl,teng23np}. First, the effects of nonuniform hoppings, such as beyond-$s$-orbital hoppings and spin-orbit couplings, deserve systematic analyses. Meanwhile, the detailed studies of interaction-driven phase diagrams may find exotic correlated phases. On the other hand, a material search with crystal-net category \cite{neves24npjcm} and corresponding synthesis can boost the discovery of relevant materials. Finally, our lattice models can be directly engineered and studied with synthetic matter, such as ultracold atoms.

\begin{acknowledgments}
The author especially thanks Joel Moore for fruitful discussions and important feedback on the manuscript. They also thank Chunxiao Liu for introducing the Bilbao Crystallographic Server \cite{bilbao11}. This work was supported by the Air Force Office of Scientific Research under Grant No. FA9550-22-1-0270. Y.-P.L. acknowledges the fellowship support from the Gordon and Betty Moore Foundation through the Emergent Phenomena in Quantum Systems (EPiQS) program.
\end{acknowledgments}




\bibliography{reference}

\clearpage
\onecolumngrid

\begin{center}{\large\bf
Supplementary information: Sublattice polarization from destructive interference on common lattices
}\end{center}

\setcounter{secnumdepth}{3}
\setcounter{equation}{0}
\setcounter{figure}{0}
\renewcommand{\theequation}{S\arabic{equation}}
\renewcommand{\thefigure}{S\arabic{figure}}
\newcommand\Scite[1]{[S\citealp{#1}]}
\makeatletter \renewcommand\@biblabel[1]{[S#1]} \makeatother


\section{Sublattice-polarized states on two- and three-dimensional lattices}

In the main text, we have shown the representative sublattice-polarized states (SLPSs) on common two- and three-dimensional (2D and 3D) lattices. In this section, we further present the SLPSs in all sublattices at nonzero high-symmetry points of the Brillouin zone (BZ). In most of these cases, the SLPSs in the sublattices $\tau$'s have staggered wavefunctions $\psi_{i\tau}=\pm1$ at different Bravais lattice sites $i$'s. This simple and commensurate structure allows for a direct identification of high-symmetry momentum $\mbf k_\text{SLPS}^\tau$ for each SLPS $(\tau,\mbf k_\text{SLPS}^\tau)$. The BZs for these lattices are also drawn, so that the SLPSs can be connected more easily to their corresponding high-symmetry points.

\begin{figure}[b]
\centering
\includegraphics[scale=1]{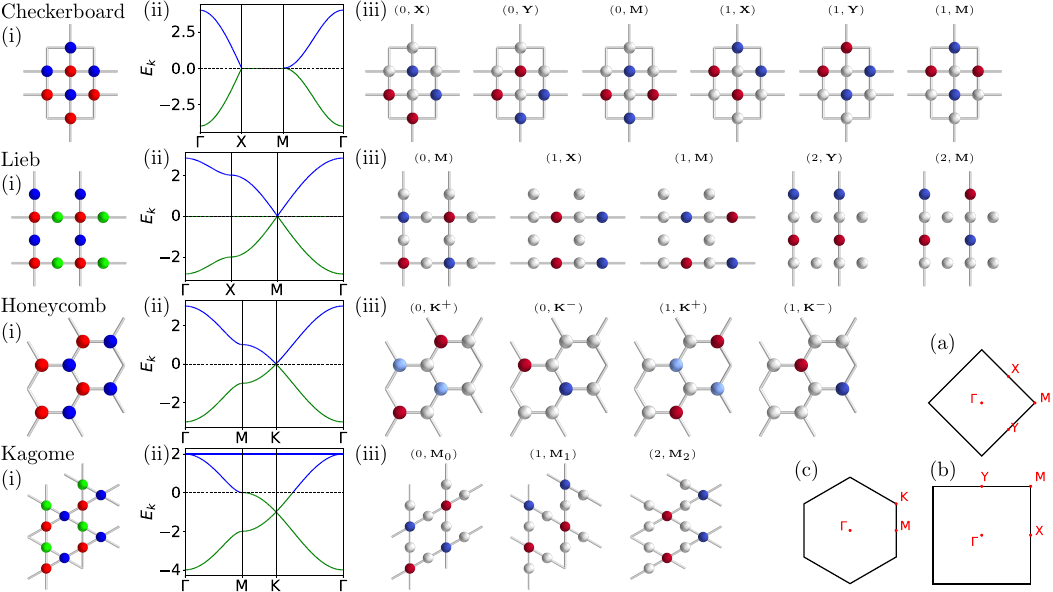}
\caption{\label{suppfig:2d} Illustration of SLPSs on the 2D lattices. For each lattice, we show (i) the lattice with sublattices marked by rainbow colors ($\tau=0,1,2,\dots$ from red to blue), (ii) the band structure with SLPS energy $\epsilon_\text{SLPS}^\tau$ indicated by dashed line, and (iii) the wavefunctions of representative SLPSs $(\tau,\mbf k^\tau_\text{SLPS})$ with $\tau$-connected bonds illustrated. The BZs for (a) checkerboard, (b) Lieb, (c) honeycomb, and kagome lattices are also presented. For the honeycomb lattice, the labels $\mbf K^\pm$ correspond to $w^3_1=(1/\sqrt{6})(2,-1,-1)$ and $w^3_2=(1/\sqrt{2})(0,1,-1)$, respectively.}
\end{figure}

We first present the results on the 2D lattices (Fig.~\ref{suppfig:2d}). For the checkerboard and Lieb lattices, the square Bravais lattice indicates a square BZ. Nonzero high-symmetry points include the two edge centers $\mbf X$ and $\mbf Y$, as well as the corner $\mbf M$. These points correspond to the $2\times1$, $1\times2$, and $\sqrt2\times\sqrt2$ periodicities, respectively. Meanwhile, the honeycomb and kagome lattices live on triangular Bravais lattice and have hexagonal BZ. Nonzero high-symmetry points include the three edge centers $\mbf M_{0,1,2}$, as well as the two corners $\mbf K$ and $\mbf K'=-\mbf K$. These points correspond to three $2\times1$ periodicities and the $\sqrt3\times\sqrt3$ periodicity, respectively.

\begin{figure}[t]
\centering
\includegraphics[scale=1]{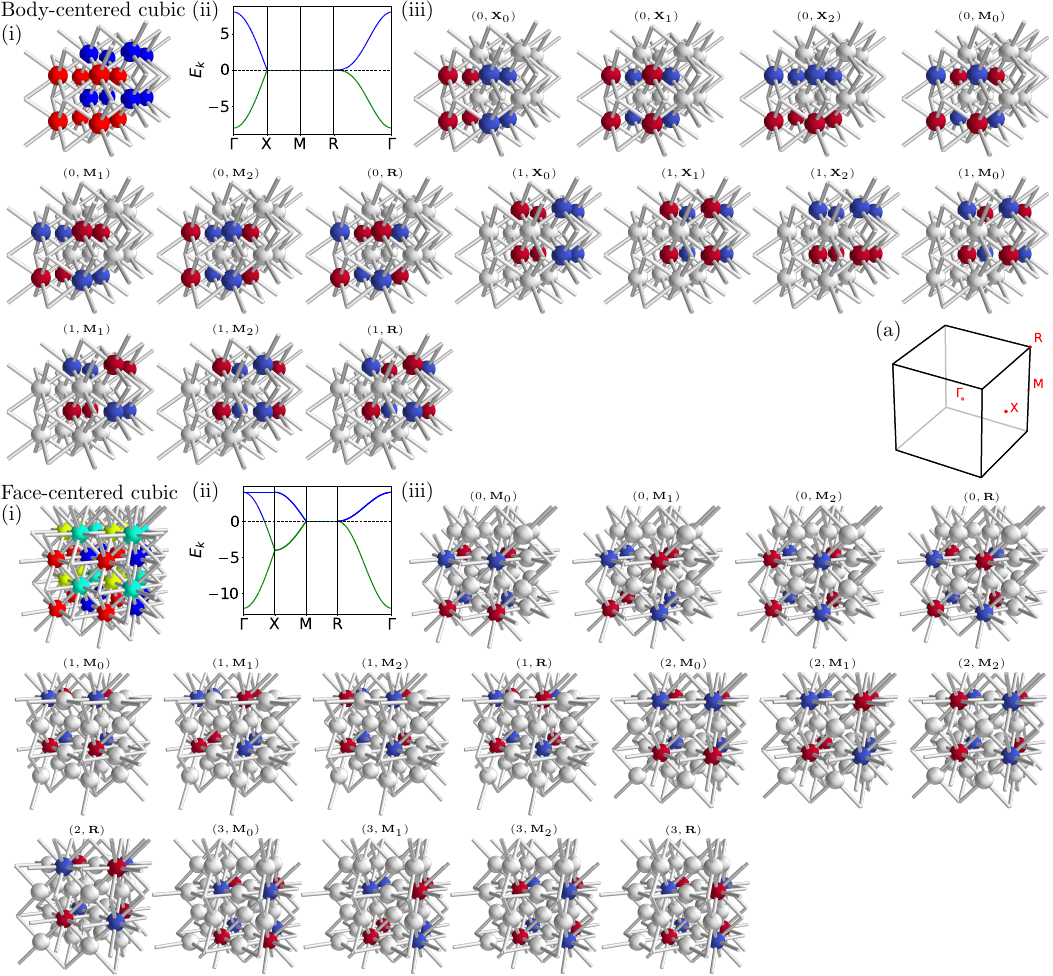}
\caption{\label{suppfig:3dsc} Illustration of SLPSs on the 3D lattices on simple cubic Bravais lattice. The angles of view are chosen for clear illustration. The BZ (a) is also shown.}
\end{figure}

\begin{figure}[t]
\centering
\includegraphics[scale=1]{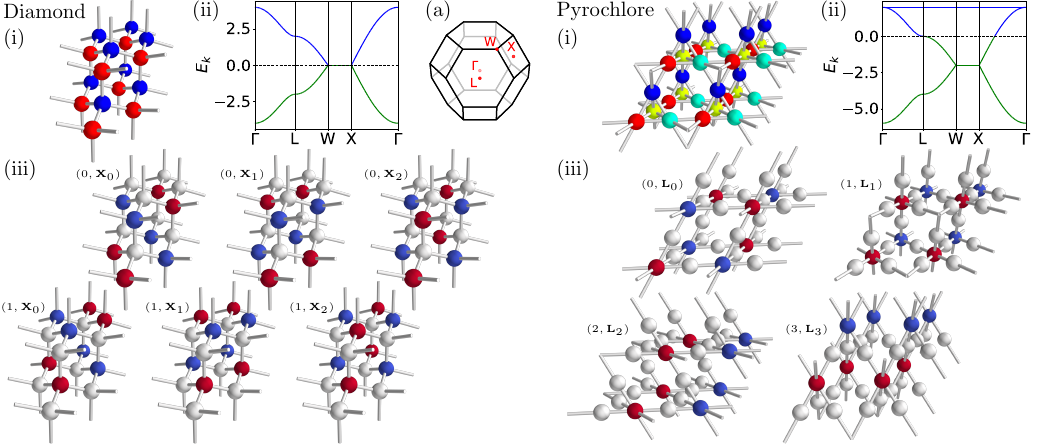}
\caption{\label{suppfig:3dfcc} Illustration of SLPSs on the 3D lattices on FCC Bravais lattice. The angles of view are chosen for clear illustration. The BZ (a) is also shown.}
\end{figure}

We next present the results on the 3D lattices. For the body-centered cubic (BCC) and face-centered cubic (FCC) lattices (Fig.~\ref{suppfig:3dsc}), the simple cubic Bravais lattice indicates a simple cubic BZ. Nonzero high-symmetry points include the three face centers $\mbf X_{0,1,2}$, the three edge centers $\mbf M_{0,1,2}$, and the corner $\mbf R$. These points correspond to three $2\times1\times1$ periodicities, three $\sqrt2\times\sqrt2\times1$ periodicities, and the $\sqrt2\times\sqrt2\times\sqrt2$ periodicity, respectively. Meanwhile, the diamond and pyrochlore lattices (Fig.~\ref{suppfig:3dfcc}) live on FCC Bravais lattice and have BCC BZ. Nonzero high-symmetry points include the four hexagon-face centers $\mbf L_{0,1,2,3}$ and the three square-face centers $\mbf X_{0,1,2}$. These points correspond to four $2\times1\times1$ periodicities and three $\sqrt3\times\sqrt3\times1$ periodicities, respectively. Note that the commonly discussed points also include the corners $\mbf W$'s and the edge centers, which we do not consider here.

\section{Robustness of sublattice-polarized states against further-neighbor hoppings}

In the main text, we have explained how the SLPSs remain robust against further-neighbor hoppings under point-group symmetries. Here we present an alternative proof based on the mathematical induction.

Assume that a sublattice $\tau$ has nearest-neighbor connection indices $N^{\tau\tau'}_1\geq2$ with all other sublattices $\tau'$'s. Our target is an SLPS $\psi_{i\tau}$ in $\tau$ under uniform nearest-neighbor hoppings. For simple and transparent discussions, we take the Bravais lattice in $\tau$ as the platform for our proof. For any site $i'\tau'$ in another sublattice, its position is located at the center of a unit bond, face, or volume of the Bravais lattice. The $N^{\tau\tau'}_1$ sites $i\tau$ in this destructive-interference (DI) unit are the nearest neighbors of $i'\tau'$, which obey a point-group symmetry $X^{\tau\tau'}$ at $i'\tau'$. Under uniform nearest-neighbor hoppings, the DI occurs in this DI unit, where the wavefunctions sum to zero $\sum_i^{i'\tau',1}\psi_{i\tau}=0$. Note that the sites $i'\tau'$ may not appear in all of the units with the same dimension and size as the DI units. For the 2D kagome and 3D pyrochlore lattices, $i'\tau'$ do not appear at all 1D bond centers. Meanwhile, for the 2D honeycomb and 3D diamond lattices, $i'\tau'$ appear at the centers of 2D triangles and 3D tetrahedrons only in one orientation, respectively. We consider the DI more carefully when the DI units only partially occupy the same-size units at the lattice dimension. The DI units may occupy half of the same-size units, as on the 2D honeycomb and 3D diamond lattices. Meanwhile, the other half empty units may be related by the inversion symmetry. When the SLPS carry a momentum $\mbf k_\text{SLPS}^\tau$, the wavefunctions in the empty units are given by those in the DI units at the opposite momentum $-\mbf k_\text{SLPS}^\tau$. Therefore, these empty units can host ``ghost DI'' and be considered equivalently to the DI units.

We prove that the SLPS remains robust against uniform further-neighbor hoppings at each range. Our proof considers the DI of the SLPS $\psi_{i\tau}$ to a site $i'\tau'$. On the Bravais lattice in $\tau$, the sites $i\tau$ are categorized into distinct groups by their distances to the site $i'\tau'$. Each $m$-th group involves all sites at the $n_m$-th neighbor and obeys the $X^{\tau\tau'}$ symmetry at $i'\tau'$. To show the robustness, we prove that DI is valid for each group under uniform hoppings. First, we have DI under uniform nearest-neighbor hoppings, which corresponds to a zero sum of the wavefunctions $\sum_i^{i'\tau',n_1=1}\psi_{i\tau}=0$ in the $m=1$ group. Now assume that the DI zero sum $\sum_i^{i'\tau',n_m}\psi_{i\tau}=0$ remains valid in the $m$-th group. When the $(m+1)$-th group is considered, one can find a set of DI units (including the ghost ones) which covers the whole $(m+1)$-th group and obeys the $X^{\tau\tau'}$ symmetry
\begin{equation}
0=\sum (\text{DI units})=\sum_{m'=1}^{m+1}C_{m'}\sum_i^{i'\tau',n_{m'}}\psi_{i\tau}.
\end{equation}
Note that the nonnegative integer coefficient $C_{m'}$ is uniform in each $m'$-th group, which is enforced by the $X^{\tau\tau'}$ symmetry. In particular, the coefficient $C_{m+1}$ is positive due to the coverage of the $(m+1)$-th group. It is straightforward that the zero sum is valid in the $(m+1)$-th group $\sum_i^{i'\tau',n_{m+1}}\psi_{i\tau}=0$, and DI occurs when uniform hoppings are introduced at this range. According to the mathematical induction, the SLPS remains robust against uniform further-neighbor hoppings at each range.

\section{Hartree-Fock ground states under sublattice polarization}

In the main text, we have presented the Hartree-Fock ground states under pure onsite or nearest-neighbor Hubbard repulsion $U_{0,1}>0$ at zero SLPS energy (Fig.~5). Here we discuss more details of our analysis. Our Hartree-Fock analysis adopts a spatially unrestricted formalism \Scite{lin23ax}, which supports an unbiased determination of mean-field ground states under energy minimization. Note that this formalism can capture the ground states with coexisting spin and charge orders. Our focus is on the symmetry-breaking particle-hole orders which develop at the weakest couplings. The finite-size lattices we consider have $N_0\times N_1\times N_2$ Bravais lattices under periodic boundary condition.

We first discuss the results on the 2D lattices. To secure the symmetry breaking in the thermodynamic limit, we compute the ground states at increasing lattice sizes and make sure that the ordering strength does not tend to zero. For the checkerboard, Lieb, honeycomb, and kagome lattices, our computation goes up to the $32\times32\times1$,  $24\times24\times1$, $30\times30\times1$, and $24\times24\times1$ Bravais lattices, respectively. The repulsions we use for these lattices are $(U_0,U_1)=(2,0.5)$, $(1,0.1)$, $(4,0.5)$, and $(2,0.2)$.

In addition to the descriptions in the main text, there are a few notes which we should make here. First, the fillings for these lattices at zero SLPS energy are $1/2$, $1/2$, $1/2$, and $(1/3)(1+1/4)$ fillings. Second, for the Lieb lattice under the onsite repulsion, the $\tau=0$ sublattice can also develop a ferromagnetism (FM), which is much weaker than and opposite to the FM in the $\tau=1,2$ sublattices. Meanwhile, under the nearest-neighbor repulsion, the intra-unit-cell (IUC) charge-density modulation (CDM) is $s$-wave. This order is a type of $s$-wave Pomeranchuk orders, which are known to develop large domains with different charge densities. Third, the SLP splittings open gaps only on the two-sublattice lattices, while the other lattices manifest SLP Fermi-surface splittings. The ground-state patterns are usually nonuniform in the latter metallic situation. Finally, our finite-size computation may not fully capture the effect of Fermi-surface nesting, such as on the kagome lattice, since it occurs at very low scale and requires very high resolution. Nevertheless, the ground states we find on the kagome lattice are consistent with the functional renormalization-group (FRG) results \Scite{kiesel13prl,wang13prb,profe24ax}.

We next discuss the results on the 3D lattices. Due to the limit of computational capability, it is hard to observe the trend of ordering strength with increasing lattice size as in 2D. Here we compute the ground states on the $8\times8\times8$ Bravais lattice, which is the maximal isotropic lattice with appropriate periodicity we can achieve reasonably. The repulsions we use for the BCC, FCC, diamond, and pyrochlore lattices are $(U_0,U_1)=(2,0.5)$, $(3,0.5)$, $(4,0.5)$, and $(2,0.2)$, respectively. For the two-sublattice BCC and diamond lattices, the zero SLPS energy occurs at the $1/2$ filling. Meanwhile, for the four-sublattice FCC and pyrochlore lattices, we choose the fillings $(1/4)(1+4/8)$ and $(1/4)(1+5/8)$, respectively, where the eigenstates of $8\times8\times8$ Bravais-lattice tight-binding models have zero energy. It should be noted that the fillings on the latter two lattices may only be close to but not exactly at zero energy in the thermodynamic limit. In the discretized BZs with high resolutions (for example, with $64^3$ momentum points), the chosen fillings deviate slightly from zero energy.

The ground states on the FCC and pyrochlore lattices deserve further discussion. Since the fillings are away from the $1/2$ filling, the computations on these two lattices are much harder to converge. Furthermore, the finite-size effect may lead to some features that are expected to vanish in the thermodynamic limit. For the FCC lattice under $U_0=3$, the lowest-energy states manifest some planar domains with antiferromagnetism (AFM) and IUC CDM. These planar domains are embedded in an environment without any symmetry-breaking order. Meanwhile, the FM state with secondary IUC CDM in the main text has a slightly higher energy. In the thermodynamic limit, subdimensional symmetry-breaking orders usually lose to the uniform ones energetically. Therefore, we expect the ground state to be the latter. Under $U_1=0.5$, the IUC CDM is slightly modulated and accompanied by a much weaker spin pattern. These features may result from the metallic feature and finite-size effect. On the other hand, for the pyrochlore lattice under $U_1=0.2$, the IUC CDM is accompanied by a secondary charge density wave (CDW) at an $\mbf L$. This CDW may vanish in the weak-coupling and thermodynamic limits.

Finally, we note that it will be worth verifying the ground states by other numerical methods with higher resolution. Our finite-size results have indicated the possible ground states in the thermodynamic limit. Furthermore, they can be directly engineered and examined in the synthetic-matter experiments, which are always finite-size. Nevertheless, the higher resolution computations can confirm whether our results are the true ground states in the thermodynamic limit. Feasible methods include momentum-space Hartree-Fock analysis with fixed ans\"atze, random-phase approximation (RPA), and FRG.

\end{document}